\documentclass[a4paper,11pt]{article}
\usepackage{amsmath,amssymb,amsfonts,amsthm,latexsym} 

\usepackage{pos}
\usepackage[normalem]{ulem}
\usepackage[utf8]{inputenc}
\usepackage[T1]{fontenc}
\usepackage{cancel}
\usepackage{graphicx}
\usepackage{latexsym}
\usepackage{bbm}
\usepackage{epsfig}
\usepackage{epstopdf}
\usepackage{subcaption}
\usepackage{color}
\usepackage{comment}
\usepackage{slashed}
\usepackage{placeins}
\usepackage{cleveref}
\usepackage{setspace}
\usepackage{url}
\usepackage{tabularx}
\usepackage{xcolor}
\usepackage{enumitem}
\usepackage{hyperref}
\usepackage{diagbox}
\usepackage{dcolumn}
\usepackage{xfrac}
\usepackage{pgf}

\newcommand{\dd}[1]{\mathop{\mathrm{d}#1}}

\newcommand{\Nt}{N_{\rm t}}
\newcommand{\Tc}{T_{\rm c}}

\newcommand{\tauf}{\tau_\mathrm{F}}

\newcommand{\Tr}{\textrm{Tr}\,}

\newcommand{\be}{\begin{equation}} 
\newcommand{\ee}{\end{equation}}
\def\ba#1\ea{\begin{align}#1\end{align}}
\newcommand{\bea}{\begin{eqnarray}} 
\newcommand{\eea}{\end{eqnarray}}
\def\lsim{\mathrel{\raise.3ex\hbox{$<$\kern-.75em\lower1ex\hbox{$\sim$}}}}
\def\gsim{\mathrel{\raise.3ex\hbox{$>$\kern-.75em\lower1ex\hbox{$\sim$}}}}

\newcommand{\bra}{\left\langle}
\newcommand{\ket}{\right\rangle}
\newcommand{\MEE}{M_\mathrm{EE}}

\renewenvironment{thebibliography}[1]{
  \begin{oldthebibliography}{#1}
    \setlength{\itemsep}{0em}
    \setlength{\parskip}{0em}
}
{
  \end{oldthebibliography}
}

\title{Adjoint chromoelectric and -magnetic correlators \\ with gradient flow}

\author*[a,b,1]{Viljami Leino}

\affiliation[a]{Helmholtz Institute Mainz, Staudingerweg 18, 55128 Mainz, Germany}
\affiliation[b]{Institut für Kernphysik, Johannes Gutenberg-Universität Mainz,\\
             Johann-Joachim-Becher-Weg 48, 55128 Mainz, Germany}

\emailAdd{viljami.leino@uni-mainz.de}

\abstract{
\vspace*{-6mm}
When QCD is described by a nonrelativistic effective field theory, operators consisting of gluonic correlators of two chromoelectric or -magnetic fields will often appear in descriptions of quarkonium physics. 
At zero T, these correlators give the masses of gluelumps and the moments of these correlators can be used to understand the inclusive P-wave decay of quarkonium. 
At finite T these correlators define the diffusion of the heavy quarkonium. However, these correlators come with a divergent term in lattice spacing which needs to be taken care of. 
We inspect these correlators in pure gauge theory with gradient flow smearing, which should allow us to reduce and remove the divergence more carefully. 
In these proceedings, we focus on the effect of gradient flow to these correlators and the reduction of this divergence.
}

\note{for the TUMQCD Collaboration}
\FullConference{The 40th International Symposium on Lattice Field Theory (Lattice 2023)\\
July 31st - August 4th, 2023\\
Fermi National Accelerator Laboratory\\}

\begin{document}
\maketitle

\section{Introduction}
When QCD is described by a nonrelativistic effective field theory,
terms consisting of components of field strength tensor $F_{\mu\nu}$ become building blocks for multiple interesting operators~\cite{Brambilla:2021mpo}.
Especially, the correlation function of two chromoelectic fields surfaces in many physically different situations, such as the diffusion of heavy quarks in a strongly coupled medium.
The heavy quark diffusion coefficient $\kappa$  
is defined as two chromoelectric fields connected with a fundamental Wilson line~\cite{Caron-Huot:2009ncn}.
This case, where the operators are connected with a fundamental Wilson line, has been comprehensively studied on the lattice and recently also the leading corrections in heavy quark mass
have been included. These corrections are given by a correlator of two chromomagnetic fields $B$. For some recent studies see for example~\cite{Brambilla:2020siz,Brambilla:2022xbd,Banerjee:2022uge,Banerjee:2022gen}.

Nevertheless, there is a less-studied version of the EE-correlator; the case in which the chromoelectric fields are connected with a Wilson line in the adjoint representation of the gauge group.
Originally this operator was constructed 
for the study of gluelumps~\cite{Jorysz:1987qj}, 
where its ground state describes the mass of the second lightest gluelump.
Respectively, the mass of the lightest gluelump is given by the BB-correlator.
While gluelump states themselves don't seem to exist in nature, they describe the short distance regime of static hybrids 
and the gluelump masses are required as a matching coefficients for hybrid static potentials~\cite{Berwein:2015vca}.
The full mass spectrum of gluelumps has been studied on the lattice before in Refs.~\cite{Foster:1998wu,Marsh:2013xsa,Herr:2023xwg}. 

Moreover, in potential non-relativistic QCD (pNRQCD), moments of the EE-correlation function appear: $\mathcal{E}_n \propto \displaystyle\int_0^\infty \dd t t^n \langle E(t)\Phi E(0) \rangle$, where $\phi$ is an adjoint Wilson line. 
Especially, the third moment $\mathcal{E}_3$ enters as input for the pNRQCD description of 
the inclusive annihilation rate 
of a P-wave spin-triplet quarkonium 
into light hadrons $\Gamma_{\chi_{QJ}}$~\cite{Brambilla:2001xy,Brambilla:2002nu,Brambilla:2020xod}.

Lastly, at finite temperature, the in-medium evolution of heavy quarkonium can
be described purely 
by two free parameters $\kappa$ and $\gamma$~\cite{Brambilla:2017zei}. 
$\kappa$ decodes the in-medium width of the quarkonium and describes the diffusion
of heavy quarkonium. The operator encoding the heavy quarkonium diffusion takes the form of an adjoint EE-correlation function, which separates it from the heavy quark case. 
Perturbatively the two $\kappa$'s agree at next-to-leading-order level. 
On the other hand, the parameter $\gamma$ describes the mass shift of the quarkonium and
is proportional to $\mathcal{E}_0$. Recently these correlators have been 
discussed in~\cite{Scheihing-Hitschfeld:2023tuz}.

With a wide range of possible applications,
it is clear that there is a demand for a modern non-perturbative high precision lattice calculations of these correlators.
Unfortunately, the Wilson line connecting the chromoelectric fields comes with a divergent self-energy contribution. 
In the previous studies of the heavy quark diffusion coefficient, 
the operator was defined with the chromoelectric fields inserted into a Polyakov loop and further normalized with the Polyakov loop, which divides out the divergence. 
With correlation functions consisting of finite length Wilson lines 
such normalization is not an option, since a finite Wilson line is not a gauge invariant observable.
For gluelumps the divergence manifests on the lattice  as a linear divergent mass shift. After removing the divergence, fixing the associated constant is equivalent to choosing a scheme. In previous gluelump studies~\cite{Bali:2003jq} 
the mass shift was fixed in the RS scheme. In these proceedings, 
we try to remove the divergence by using two different approaches different 
from the RS-scheme. 
This divergence problem is not dissimilar to the divergences encountered 
in lattice studies of quasi parton distribution functions~\cite{Chen:2016fxx,Monahan:2017hpu}.

\section{Construction of the correlation functions}
We study the adjoint chromoelectric and -magnetic correlators at zero and finite temperature. The correlators are given
in continuum as:
\begin{align}
    G_\mathrm{EE}(t) = \bra 0|g E^{i,a}(t,0)G_{ab}(t,0)gE^{i,b}(0,0)|0\ket\,,\label{Eq:basiccorrsEE} \\
    G_\mathrm{BB}(t) = \bra 0|g B^{i,a}(t,0)G_{ab}(t,0)gB^{i,b}(0,0)|0\ket\,.\label{Eq:basiccorrsBB}
\end{align}
The chromoelectric correlator can be calculated on the lattice:
\begin{equation}
    G_\mathrm{EE}(t) = \frac{1}{V} \sum_{t^\prime} \sum_{\vec{x}} \bra E^{i,\alpha}(t^\prime+t,\vec{x})
    G_{\alpha\beta}^\mathrm{A}(t^\prime+t,t^\prime)E^{i,\beta}(t^\prime,\vec{x})\ket\,,
\end{equation}
where $G_{\alpha\beta}^\mathrm{A}$ is an adjoint temporal Wilson line and the chromoelectric operators are defined with the clover discretization.
The finite discretization of the field strength tensor components come with a lattice specific multiplicative renormalization due to gluonic self-energy 
diagrams inside the discretized fields~\cite{Lepage:1992xa}. 
Additionally, the chromomagnetic fields have a finite anomalous dimension, 
see e.g.~\cite{Banerjee:2022uge,Brambilla:2023vwm}, which has to be accounted for when the zero flow time limit is taken.

We use gradient flow algorithm~\cite{Luscher:2009eq} to both reduce the noise via smearing and also to
renormalize the field strength tensor components. 
We have recently shown that the chromoelectric fields are non-perturbatively renormalized under the gradient flow~\cite{Brambilla:2023fsi}.
This renormalization property applies also to chromomagnetic fields. 
For the gradient flow to fully renormalize the field strength tensor components, 
the smearing radius $\sqrt{8\tauf}$ has to be larger than one lattice spacing $a$. 
Meanwhile, we also want to avoid contact terms that will appear if the smearing radius grows too large compared to the separation $t$
between the operators. 
Hence, we limit the analysis to a regime:
\begin{equation}\label{eq:flowtimelimitcrit}
    1 < \frac{\sqrt{8\tauf}}{a} < \begin{cases}\frac{t-2}{2} & \text{for the EE-correlator} \\ \;\, \frac{t}{2} & \text{for the BB-correlator} \end{cases}\,,
\end{equation}
where we have separate maximum flow times for both chromoelectric and -magnetic fields. 
This difference comes from the orientation of the respective field strength tensor components. Chromoelectric fields are parallel with the connecting adjoint Wilson line, while the chromomagnetic fields are perpendicular to it.
At large separations $t>\Nt/2$, we also make sure to avoid contact terms over the periodic boundaries: $2\sqrt{8\tauf}<\Nt-t$.
The adjoint operators, and the adjoint Wilson line are related to their fundamental counterparts as 
\begin{alignat}{2}
    E^{i,\alpha} &= \Tr(\lambda^\alpha E^i)&
    \quad B^{i,\alpha} &= \Tr(\lambda^\alpha B^i)\\
    G_{\alpha \beta}(t,0) &= \prod_{\tau=0}^{t-1} U^8_4(\tau) &
    U^8_{ij} &= \frac{1}{2}\Tr\left[ \lambda_i U^3 \lambda_j U^{3\dagger}\right]\label{eq:linkrels}\,,
\end{alignat}
where $U^n$ is a gauge link in n-dimensional representation. We use these fundamental representation definitions to implement the EE- and BB-correlators on the lattice.

\begin{table*}[t]
           \centering
           \captionsetup[subtable]{position = below}
           \hspace{-3em}
           \begin{subtable}{0.4\linewidth}
               \centering
    \begin{tabular}{c|c|c|c|c|c}
        $N_S$ & $N_T$ & $\beta$ & $a$ [fm] & $T/\Tc$ & $N_\mathrm{conf}$ \\\hline
        $20^3$ & $6$ & 6.284 & 0.060 & 1.848 & 764 \\\hline
        $20^3$ & $8$ & 6.284 & 0.060 & 1.386 & 620 \\\hline
        $40^3$ & $6$ & 6.816 & 0.030 & 3.765 & 226 \\\hline
    \end{tabular}
    \caption{finite T ensembles}
    \label{tab:lattice_parameters_finitet}
           \end{subtable}%
           \hspace*{5em}
           \begin{subtable}{0.4\linewidth}
               \centering
    \begin{tabular}{c|c|c|c|c|c}
        $N_S$ & $N_T$ & $\beta$ & $a$ [fm] & $T/\Tc$ & $N_\mathrm{conf}$ \\\hline
        $20^3$ & $40$ & 6.284 & 0.060 & 0.277 & 6000 \\\hline
        $26^3$ & $56$ & 6.481 & 0.046 & 0.261 & 6000 \\\hline
        $30^3$ & $60$ & 6.594 & 0.040 & 0.283 & 6000 \\\hline
        $40^3$ & $80$ & 6.816 & 0.030 & 0.282 & 3300 \\\hline
    \end{tabular}
    \caption{Zero T ensembles}
    \label{tab:lattice_parameters_zerot}
           \end{subtable}
    \caption{Parameters for our current lattice ensembles} 
           \label{table_fulltable1}
       \end{table*}
We have generated a set of ensembles\footnote{The zero T ensembles were originally generated for~\cite{Brambilla:2023fsi}} 
both at finite and zero temperature to study these correlators. The current status of the simulations together with the
exact simulation parameters 
is presented in the table~\ref{table_fulltable1}, 
where $a$ in fm units is based on the scale setting from~\cite{Necco:2001xg} with assumption $r_0=0.5$fm, 
and $T/\Tc$ is based on scale setting from~\cite{Francis:2015lha}.
We produce the lattice configurations with Wilson action using the heatbath and overrelaxation algorithms. 
The gradient flow is evolved using the solvers described in~\cite{Bazavov:2021pik}.

\section{Removal of the Wilson line self-energy}
A Wilson line in any representation has an associated self-energy which diverges as $\propto 1/a$. 
The gradient flow will regularize this divergence by smearing out the UV-scale $1/a$ and change the divergence to $\propto 1/\sqrt{8\tauf}$.
For a Wilson line in D-dimensional representation, we will denote the self-energy as $\delta m_D$;
particularly the self-energy of an adjoint Wilson line is given by $\delta m_8$. 
The divergence $\delta m_8$
modifies the correlator by a multiplicative factor $\exp(-\delta m_8 t)$. 
and is independent of the temperature. 
In order to achieve a physical measurement of the correlators of interest:~(\ref{Eq:basiccorrsEE},\ref{Eq:basiccorrsBB}), 
the self-energy divergence has to be removed.

\begin{figure}
    \includegraphics[width=0.5\textwidth,page=25]{eff_mass_all_cors_all_taus_40.pdf}
    \includegraphics[width=0.5\textwidth]{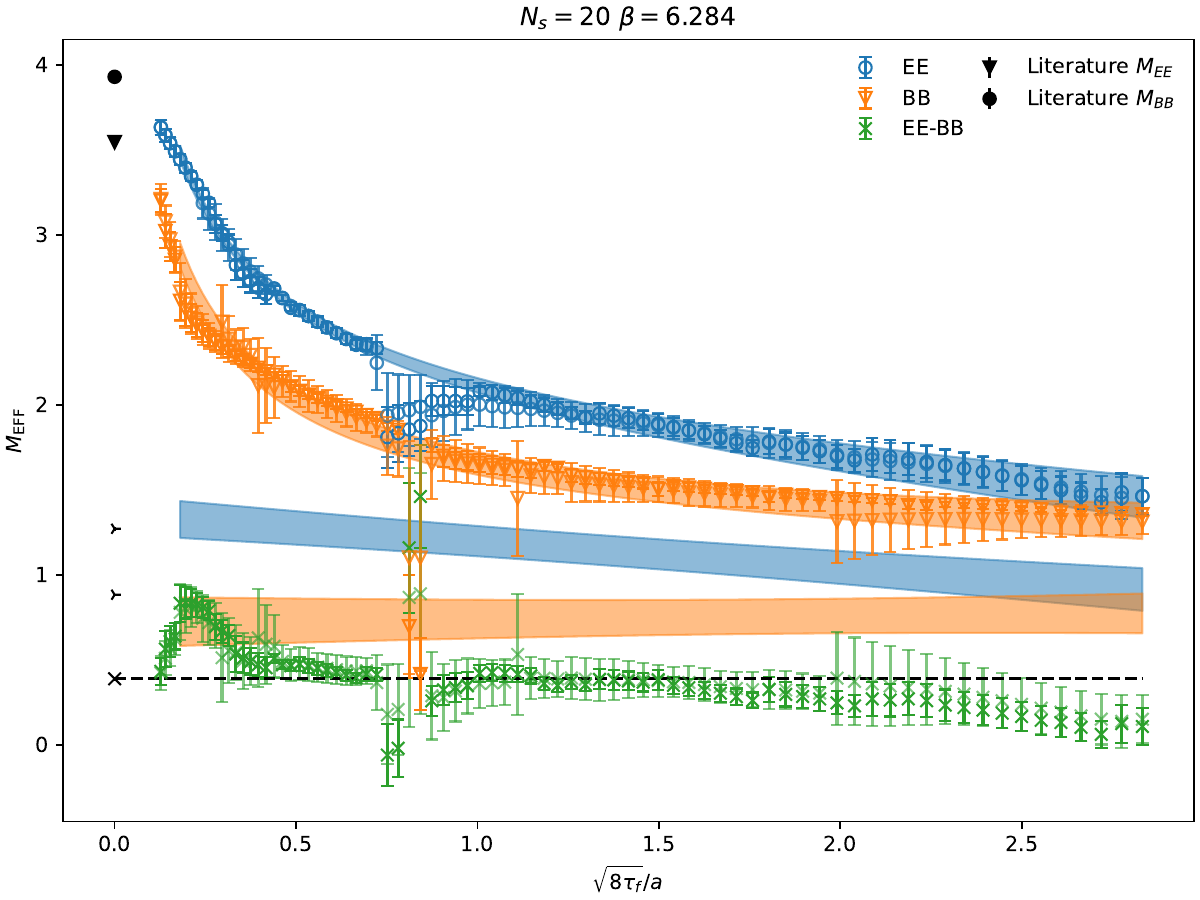}
    \caption{Ground state masses of the EE- and BB-correlation functions. Left: An example of the effective mass and respective plateau extraction. The vertical lines indicate the limits imposed by Eq.~\eqref{eq:flowtimelimitcrit}, 
    while the horizontal band shows the AIC average. The dashed line corresponds to the best individual fit. Right: The removal of the linear divergence by a fit. The black points are the recent perturbatively renormalized results from~\cite{Herr:2023xwg} before and
    after the renormalization. The shaded bands correspond to our divergence removal.
    }
    \label{fig:effectivemasses}
\end{figure}
We test the divergence removal by applying it to the ground state masses of the lowest laying gluelump states. Gluelumps are chosen as the test operator for two reasons.
Firstly, there exist a recent perturbatively renormalized gluelump study~\cite{Herr:2023xwg} that uses same or similar lattice spacings as we do, which allows for direct comparison.
Secondly, the ground state mass decays exponentially $G_\mathrm{EE} = A\exp(-\MEE(a,\tauf)t)$.
Ergo, the divergence becomes a simple shift in the mass and the renormalized gluelump mass can be written as:
\begin{equation}
    \MEE^{\rm r} = \MEE(a,\tauf) - \delta m_8(a,\tauf)\,,
\end{equation}

On the left side of Fig.~\ref{fig:effectivemasses}, we plot an example of the effective masses for the correlators~(\ref{Eq:basiccorrsEE},\ref{Eq:basiccorrsBB}) 
together with a constant fit to extract the ground state masses. 
We use Akaike information criterion (AIC)~\cite{Jay:2020jkz} to find the optimal plateau at each flow time. 

\subsection{Fitting the divergence}
With the ground state masses available at each flow time, we can 
fit the divergence $\delta m_8\sim 1/\sqrt{8\tauf}$ and subtract it. This extraction is shown on the right side of Fig.~\ref{fig:effectivemasses}.
We see that the divergence extracted results, shown as shaded bands, have only 
linear flow time dependence and in theory allow for zero flow time extrapolation.
We showcase the smallest volume $L20$ only, since this allows direct comparison to the other divergence removal method that we test in the next section using the finite temperature ensembles.  
Also, the lattice parameters of the $L20$-ensemble are shared with Ref.~\cite{Herr:2023xwg}, 
which allows comparison to recent literature at finite lattice spacing.
The gluelump masses before and after the renormalization with the RS-scheme from Ref.~\cite{Herr:2023xwg} are shown in black. 
We observe that this method seems promising on replicating recent result from the literature. 
The literature results are in RS-scheme, while ours are not. However, the scheme dependent contributions to the divergence start at order of $\mathcal{O}(\alpha^2)$, 
hence the results are expected to agree at one-loop level. 
Alongside with the divergence removal,
the right hand side of Fig.~\ref{fig:effectivemasses} also shows the mass splitting between the two lowest laying gluelumps. The mass splittings between gluelumps are free of divergences. 
We observe our mass splitting to also agree well with the literature value.  

\subsection{Renormalized Polyakov loops}
Our second approach towards removing the self-energy of adjoint Wilson lines, is to use the renormalized Polyakov loops from~\cite{Gupta:2007ax}.
The local Polyakov loop $L(\vec{x})$ is defined through the thermal Wilson line, $P(x)$ as,
\begin{equation}
    L_D(x) = \Tr P(x) = \Tr \prod_{i=0}^{\Nt} U^D_{(\vec{x},i),4}\,.
\end{equation}
Using Eq.~\eqref{eq:linkrels}, the adjoint Polyakov loop can be evaluated as:
\begin{equation}\label{eq:casimir}
    L_8(\vec{x}) = \left| L_3(\vec{x})\right|^2 -1\,.
\end{equation}
Furthermore, the Polyakov loops in fundamental and adjoint representations are related trough a Casimir scaling:
\begin{equation}
    L_8(T) = L_3(T)^{d_8} = L_3(T)^{9/4}\,,
\end{equation}
where $d_D = C_2(D)/C_2(3)$ and $C_2(D)=\Tr\sum_a \lambda^a\lambda^a$ is the quadratic Casimir operator in representation $D$.
Since the Polyakov loop is an order parameter of the QCD, it is non-zero only at finite T. 

\begin{figure}
    \includegraphics[width=0.5\textwidth]{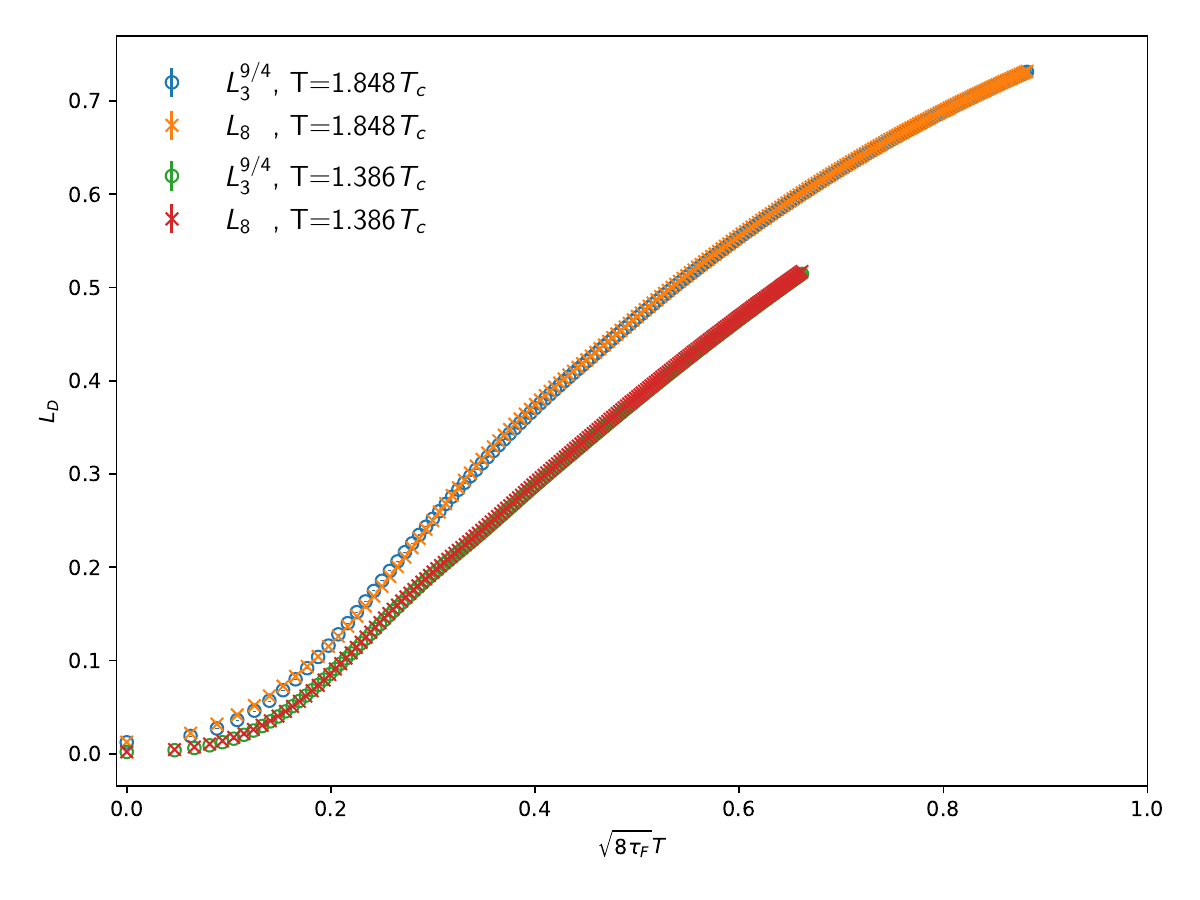}
    \includegraphics[width=0.5\textwidth]{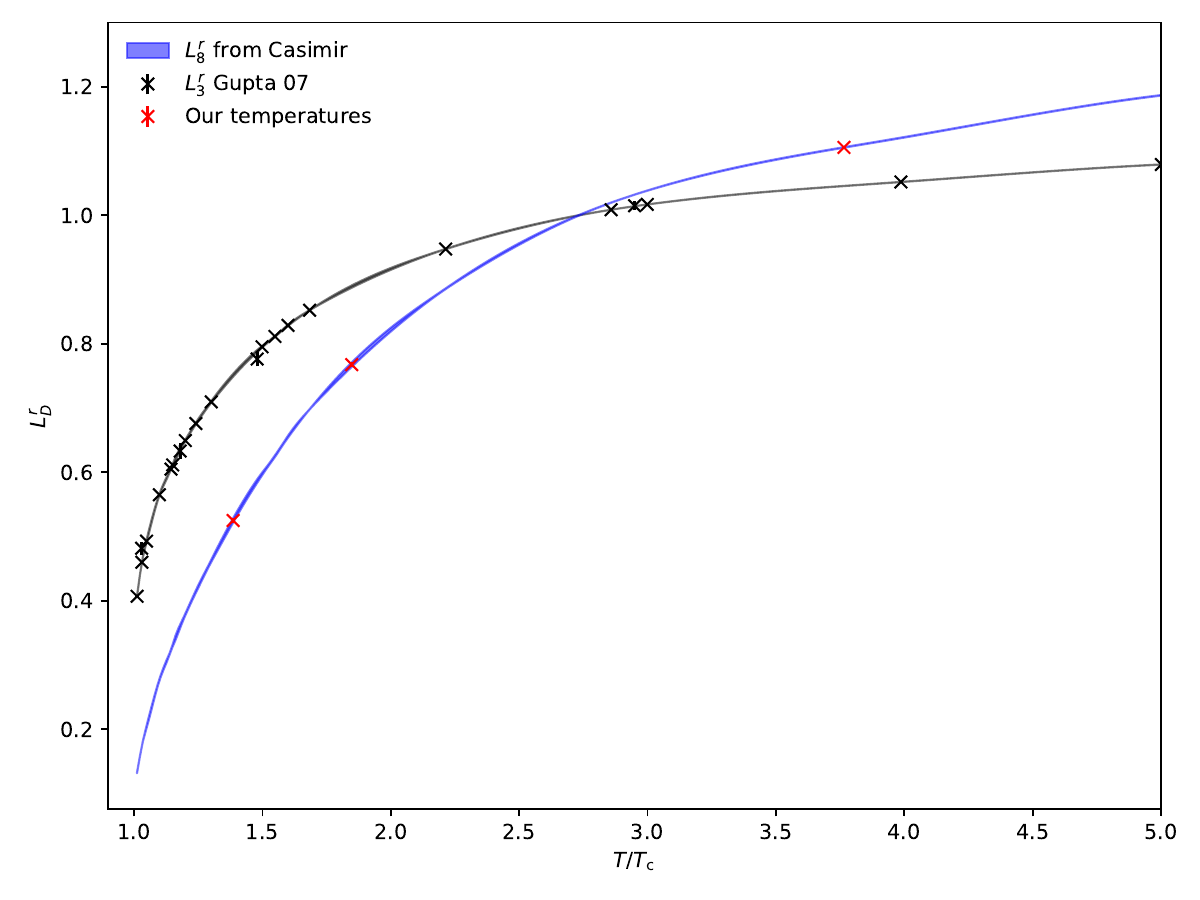}
    \caption{The Polyakov loop measurements. Left: the adjoint Polyakov loop $L_8$ compared to Casimir scaled fundamental Polyakov loops $L_3$. 
		      Right: the procedure to determine the renormalized adjoint Polyakov loop $L^\mathrm{r}_8$ from the information given in Ref.~\cite{Gupta:2007ax}.
             }
    \label{fig:Polyakovs}
\end{figure}
We measure the Polyakov loops at both adjoint and fundamental representations. On the left side of Fig.~\ref{fig:Polyakovs},
we show the Polyakov loops in both representations with $L_3$ scaled with Eq.~\eqref{eq:casimir} and observe a perfect Casimir scaling.
Now, to extract the adjoint Wilson line self-energy, we will take the renormalized fundamental Polyakov loops $L^\mathrm{r}_3$ from~\cite{Gupta:2007ax} and
use Casimir scaling to translate these to renormalizes adjoint Polyakov loops $L^\mathrm{r}_8$ at the temperatures listed in table~\ref{table_fulltable1}.
This procedure is shown on the right side of the Fig.~\ref{fig:Polyakovs}. 
The renormalized Polyakov loops are free of the self-energy divergence, 
while the non-renormalized Polyakov loops have a divergence $\sim \exp(-\delta m_D \Nt)$. 
Hence, the self-energy can be extracted as:
\begin{equation}\label{eq:howtorenormalizepolyakovloops}
    -T\left(\ln[L^b_D(T,a,\tauf)] - \ln[L^{\rm r}_D(T)]\right) = \delta m_D(a,\tauf)\,,
\end{equation}
where $T=1/\Nt$. This result should be independent of the temperature $T$ for given lattice spacing $a$ and flow time $\tauf$.
\begin{figure}
    \centering
    \includegraphics[width=0.5\textwidth,page=2]{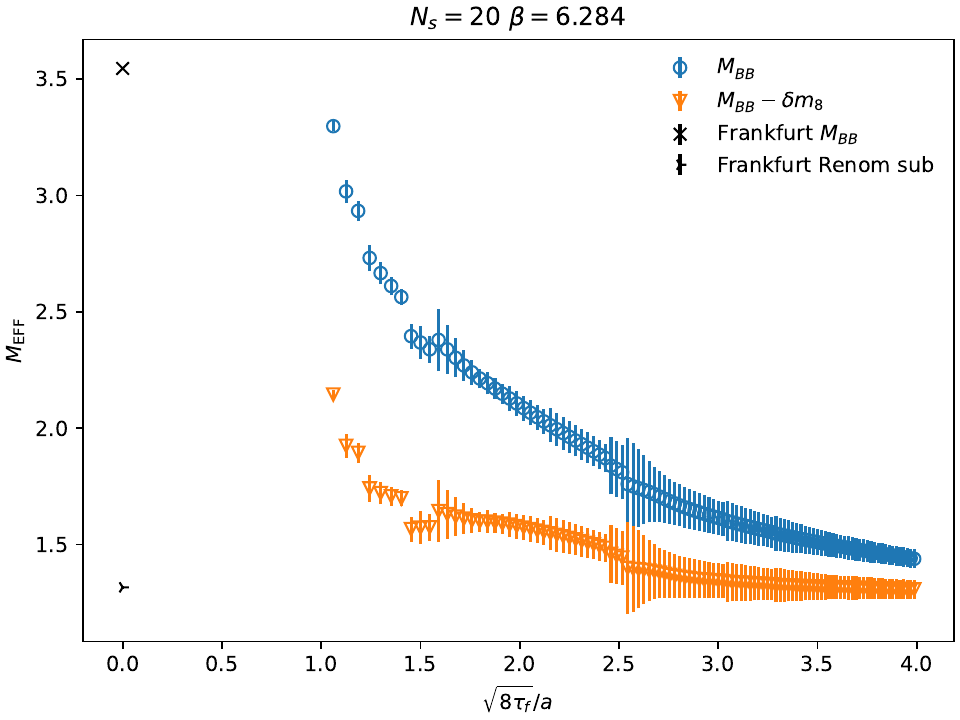}
    \caption{The lightest gluelump mass with and without the Polyakov loop based renormalization. Literature results shown in black.
             }
    \label{fig:Polyakovs2}
\end{figure}

We show the effects of this method in Fig.~\ref{fig:Polyakovs2}. 
We observe that this method has some remaining divergent shape and in general doesn't seem to work quite as well as the direct fit approach. 

\section{Conclusions}
The chromoelectric and -magnetic correlators are interesting operators to study on the lattice. 
In this proceedings we have showcased our current status of removal of the linear divergence from these operators. 
We learn that the direct fit of the $1/\sqrt{8\tauf}$ seems to work better than the Polyakov loop based renormalization. 
The next steps will be measuring all the objects mentioned in the introduction and calculating the numerical integrals for the moments $\mathcal{E}_n$.

\acknowledgments{
We acknowlede the contributions of our collaborators: N.~Brambilla, S.~Datta, J.~Mayer-Steudte, P.~Petrezky,  A.~Shindler and A.~Vairo. We also acknowledge collaboration witrh Z.~Kordov in a very initial stage of this work.
The lattice QCD calculations have been performed using the publicly available \href{https://web.physics.utah.edu/~detar/milc/milcv7.html}{MILC code}. 
The simulations were carried out on the computing facilities of the Computational Center for Particle and Astrophysics (C2PAP) in the project 'Calculation of finite T QCD correlators' (pr83pu) 
and of the SuperMUC cluster at the Leibniz-Rechenzentrum (LRZ) in the project 'Static force and other operators with field strength tensor insertions' (pn49lo). 
This research was funded by the Deutsche Forschungsgemeinschaft (DFG, German Research Foundation) cluster of excellence “ORIGINS” (\href{www.origins-cluster.de}{www.origins-cluster.de}) under Germany’s Excellence Strategy EXC-2094-390783311.}

\bibliographystyle{jhep_modified}
\bibliography{kappa.bib}

\providecommand{\href}[2]{#2}\begingroup\raggedright\begin{thebibliography}{10}

\bibitem{Brambilla:2021mpo}
N.~Brambilla, \emph{{Effective Field Theories and Lattice QCD for the X Y Z
  frontier}}, \href{https://doi.org/10.22323/1.396.0020}{\emph{PoS} {\bfseries
  LATTICE2021} (2022) 020} [\href{https://arxiv.org/abs/2111.10788}{{\ttfamily
  2111.10788}}].
\bibitem{Caron-Huot:2009ncn}
S.~Caron-Huot, M.~Laine and G.~D. Moore, \emph{{A Way to estimate the heavy
  quark thermalization rate from the lattice}},
  \href{https://doi.org/10.1088/1126-6708/2009/04/053}{\emph{JHEP} {\bfseries
  04} (2009) 053} [\href{https://arxiv.org/abs/0901.1195}{{\ttfamily
  0901.1195}}].
\bibitem{Brambilla:2020siz}
N.~Brambilla, V.~Leino, P.~Petreczky and A.~Vairo, \emph{{Lattice QCD
  constraints on the heavy quark diffusion coefficient}},
  \href{https://doi.org/10.1103/PhysRevD.102.074503}{\emph{Phys. Rev. D}
  {\bfseries 102} (2020) 074503}
  [\href{https://arxiv.org/abs/2007.10078}{{\ttfamily 2007.10078}}].
\bibitem{Brambilla:2022xbd}
N.~Brambilla, V.~Leino, J.~Mayer-Steudte and P.~Petreczky, \emph{{Heavy quark
  diffusion coefficient with gradient flow}},
  \href{https://doi.org/10.1103/PhysRevD.107.054508}{\emph{Phys. Rev. D}
  {\bfseries 107} (2023) 054508}
  [\href{https://arxiv.org/abs/2206.02861}{{\ttfamily 2206.02861}}].
\bibitem{Banerjee:2022uge}
D.~Banerjee, S.~Datta and M.~Laine, \emph{{Lattice study of a magnetic
  contribution to heavy quark momentum diffusion}},
  \href{https://doi.org/10.1007/JHEP08(2022)128}{\emph{JHEP} {\bfseries 08}
  (2022) 128} [\href{https://arxiv.org/abs/2204.14075}{{\ttfamily
  2204.14075}}].
\bibitem{Banerjee:2022gen}
D.~Banerjee, R.~Gavai, S.~Datta and P.~Majumdar, \emph{{Temperature dependence
  of the static quark diffusion coefficient}},
  \href{https://doi.org/https://doi.org/10.1016/j.nuclphysa.2023.122721}{\emph{Nuclear
  Physics A} {\bfseries 1038} (2023) 122721}
  [\href{https://arxiv.org/abs/2206.15471}{{\ttfamily 2206.15471}}].
\bibitem{Jorysz:1987qj}
I.~H. Jorysz and C.~Michael, \emph{{The Field Configurations of a Static
  Adjoint Source in SU(2) Lattice Gauge Theory}},
  \href{https://doi.org/10.1016/0550-3213(88)90210-6}{\emph{Nucl. Phys. B}
  {\bfseries 302} (1988) 448}.
\bibitem{Berwein:2015vca}
M.~Berwein, N.~Brambilla, J.~Tarr\'us~Castell\`a and A.~Vairo,
  \emph{{Quarkonium Hybrids with Nonrelativistic Effective Field Theories}},
  \href{https://doi.org/10.1103/PhysRevD.92.114019}{\emph{Phys. Rev. D}
  {\bfseries 92} (2015) 114019}
  [\href{https://arxiv.org/abs/1510.04299}{{\ttfamily 1510.04299}}].
\bibitem{Foster:1998wu}
{\scshape UKQCD} collaboration, M.~Foster and C.~Michael, \emph{{Hadrons with a
  heavy color adjoint particle}},
  \href{https://doi.org/10.1103/PhysRevD.59.094509}{\emph{Phys. Rev. D}
  {\bfseries 59} (1999) 094509}
  [\href{https://arxiv.org/abs/hep-lat/9811010}{{\ttfamily hep-lat/9811010}}].
\bibitem{Marsh:2013xsa}
K.~Marsh and R.~Lewis, \emph{{A lattice QCD study of generalized gluelumps}},
  \href{https://doi.org/10.1103/PhysRevD.89.014502}{\emph{Phys. Rev. D}
  {\bfseries 89} (2014) 014502}
  [\href{https://arxiv.org/abs/1309.1627}{{\ttfamily 1309.1627}}].
\bibitem{Herr:2023xwg}
J.~Herr, C.~Schlosser and M.~Wagner, \emph{{Gluelump masses and mass splittings
  from SU(3) lattice gauge theory}},
  \href{https://arxiv.org/abs/2306.09902}{{\ttfamily 2306.09902}}.
\bibitem{Brambilla:2001xy}
N.~Brambilla, D.~Eiras, A.~Pineda, J.~Soto and A.~Vairo, \emph{{New predictions
  for inclusive heavy quarkonium P wave decays}},
  \href{https://doi.org/10.1103/PhysRevLett.88.012003}{\emph{Phys. Rev. Lett.}
  {\bfseries 88} (2002) 012003}
  [\href{https://arxiv.org/abs/hep-ph/0109130}{{\ttfamily hep-ph/0109130}}].
\bibitem{Brambilla:2002nu}
N.~Brambilla, D.~Eiras, A.~Pineda, J.~Soto and A.~Vairo, \emph{{Inclusive
  decays of heavy quarkonium to light particles}},
  \href{https://doi.org/10.1103/PhysRevD.67.034018}{\emph{Phys. Rev. D}
  {\bfseries 67} (2003) 034018}
  [\href{https://arxiv.org/abs/hep-ph/0208019}{{\ttfamily hep-ph/0208019}}].
\bibitem{Brambilla:2020xod}
N.~Brambilla, H.~S. Chung, D.~M\"uller and A.~Vairo, \emph{{Decay and
  electromagnetic production of strongly coupled quarkonia in pNRQCD}},
  \href{https://doi.org/10.1007/JHEP04(2020)095}{\emph{JHEP} {\bfseries 04}
  (2020) 095} [\href{https://arxiv.org/abs/2002.07462}{{\ttfamily
  2002.07462}}].
\bibitem{Brambilla:2017zei}
N.~Brambilla, M.~A. Escobedo, J.~Soto and A.~Vairo, \emph{{Heavy quarkonium
  suppression in a fireball}},
  \href{https://doi.org/10.1103/PhysRevD.97.074009}{\emph{Phys. Rev.}
  {\bfseries D97} (2018) 074009}
  [\href{https://arxiv.org/abs/1711.04515}{{\ttfamily 1711.04515}}].
\bibitem{Scheihing-Hitschfeld:2023tuz}
B.~Scheihing-Hitschfeld and X.~Yao, \emph{{Real time quarkonium transport
  coefficients in open quantum systems from Euclidean QCD}},
  \href{https://doi.org/10.1103/PhysRevD.108.054024}{\emph{Phys. Rev. D}
  {\bfseries 108} (2023) 054024}
  [\href{https://arxiv.org/abs/2306.13127}{{\ttfamily 2306.13127}}].
\bibitem{Bali:2003jq}
G.~S. Bali and A.~Pineda, \emph{{QCD phenomenology of static sources and
  gluonic excitations at short distances}},
  \href{https://doi.org/10.1103/PhysRevD.69.094001}{\emph{Phys. Rev. D}
  {\bfseries 69} (2004) 094001}
  [\href{https://arxiv.org/abs/hep-ph/0310130}{{\ttfamily hep-ph/0310130}}].
\bibitem{Chen:2016fxx}
J.-W. Chen, X.~Ji and J.-H. Zhang, \emph{{Improved quasi parton distribution
  through Wilson line renormalization}},
  \href{https://doi.org/10.1016/j.nuclphysb.2016.12.004}{\emph{Nucl. Phys. B}
  {\bfseries 915} (2017) 1} [\href{https://arxiv.org/abs/1609.08102}{{\ttfamily
  1609.08102}}].
\bibitem{Monahan:2017hpu}
C.~Monahan, \emph{{Smeared quasidistributions in perturbation theory}},
  \href{https://doi.org/10.1103/PhysRevD.97.054507}{\emph{Phys. Rev. D}
  {\bfseries 97} (2018) 054507}
  [\href{https://arxiv.org/abs/1710.04607}{{\ttfamily 1710.04607}}].
\bibitem{Lepage:1992xa}
G.~P. Lepage and P.~B. Mackenzie, \emph{{On the viability of lattice
  perturbation theory}},
  \href{https://doi.org/10.1103/PhysRevD.48.2250}{\emph{Phys. Rev. D}
  {\bfseries 48} (1993) 2250}
  [\href{https://arxiv.org/abs/hep-lat/9209022}{{\ttfamily hep-lat/9209022}}].
\bibitem{Brambilla:2023vwm}
N.~Brambilla and X.-P. Wang, \emph{{Off-lightcone Wilson-line operators in
  gradient flow}},  \href{https://arxiv.org/abs/2312.05032}{{\ttfamily
  2312.05032}}.
\bibitem{Luscher:2009eq}
M.~Lüscher, \emph{{Trivializing maps, the Wilson flow and the HMC algorithm}},
  \href{https://doi.org/10.1007/s00220-009-0953-7}{\emph{Commun. Math. Phys.}
  {\bfseries 293} (2010) 899}
  [\href{https://arxiv.org/abs/0907.5491}{{\ttfamily 0907.5491}}].
\bibitem{Brambilla:2023fsi}
N.~Brambilla, V.~Leino, J.~Mayer-Steudte and A.~Vairo, \emph{{The static force
  from generalized Wilson loops on the lattice using gradient flow}},
  \href{https://arxiv.org/abs/2312.17231}{{\ttfamily 2312.17231}}.
\bibitem{Necco:2001xg}
S.~Necco and R.~Sommer, \emph{{The N(f) = 0 heavy quark potential from short to
  intermediate distances}},
  \href{https://doi.org/10.1016/S0550-3213(01)00582-X}{\emph{Nucl. Phys. B}
  {\bfseries 622} (2002) 328}
  [\href{https://arxiv.org/abs/hep-lat/0108008}{{\ttfamily hep-lat/0108008}}].
\bibitem{Francis:2015lha}
A.~Francis, O.~Kaczmarek, M.~Laine, T.~Neuhaus and H.~Ohno, \emph{{Critical
  point and scale setting in SU(3) plasma: An update}},
  \href{https://doi.org/10.1103/PhysRevD.91.096002}{\emph{Phys. Rev.}
  {\bfseries D91} (2015) 096002}
  [\href{https://arxiv.org/abs/1503.05652}{{\ttfamily 1503.05652}}].
\bibitem{Bazavov:2021pik}
A.~Bazavov and T.~Chuna, \emph{{Efficient integration of gradient flow in
  lattice gauge theory and properties of low-storage commutator-free Lie group
  methods}},  \href{https://arxiv.org/abs/2101.05320}{{\ttfamily 2101.05320}}.
\bibitem{Jay:2020jkz}
W.~I. Jay and E.~T. Neil, \emph{{Bayesian model averaging for analysis of
  lattice field theory results}},
  \href{https://doi.org/10.1103/PhysRevD.103.114502}{\emph{Phys. Rev. D}
  {\bfseries 103} (2021) 114502}
  [\href{https://arxiv.org/abs/2008.01069}{{\ttfamily 2008.01069}}].
\bibitem{Gupta:2007ax}
S.~Gupta, K.~Huebner and O.~Kaczmarek, \emph{{Renormalized Polyakov loops in
  many representations}},
  \href{https://doi.org/10.1103/PhysRevD.77.034503}{\emph{Phys. Rev. D}
  {\bfseries 77} (2008) 034503}
  [\href{https://arxiv.org/abs/0711.2251}{{\ttfamily 0711.2251}}].
\end{thebibliography}\endgroup

\end{document}